\documentclass[article,twocolumn,showpacs,preprintnumbers,amsmath,amssymb]{revtex4}
\usepackage[dvips]{color}
\usepackage{rotating}
\usepackage{graphicx}
\usepackage{dcolumn}
\usepackage{bm}
\usepackage{epsfig}
\usepackage{hyperref}
\usepackage{ulem}

\newcommand{\lton}{\mathrel{\lower.9ex
\hbox{$\stackrel{\displaystyle <}{\sim}$}}}
\newcommand{\gton}{\mathrel{\lower.9ex
\hbox{$\stackrel{\displaystyle >}{\sim}$}}}

\newcommand{\shk}[1]{{k}\!\!\!/}
\newcommand{\shq}[1]{{q}\!\!\!/}

\newcommand{\barQ}{{\bar{Q}}}

\newcommand{\bfr}{{\bf{r}}}
\newcommand{\hatr}{{\hat{r}}}

\newcommand{\calO}{{\cal{O}}}

\begin{document}

\begin{flushright}
\vskip .5cm
\end{flushright} \vspace{1cm}

\title{Collisional and thermal dissociation of $J/\psi$ and $\Upsilon$ states at the LHC }

\author{Samuel Aronson$^1$}%
\email{samuelaronson@umail.ucsb.edu}

\author{Evan Borras$^1$}%
\email{evanborras@umail.ucsb.edu}

\author{Brunel Odegard$^1$}%
\email{bcodegard@umail.ucsb.edu}

\author{Rishi Sharma$^2$}%
\email{rishi@theory.tifr.res.in}

\author{Ivan Vitev$^3$}
\email{ivitev@lanl.gov}

\affiliation{$^1$ University of California, Santa Barbara, College of Creative Studies, Isla Vista, CA 93106, USA}   
\affiliation{$^2$ Tata Institute of Fundamental Research,  Mumbai, Maharashtra 400005, India}
\affiliation{$^3$ Los Alamos National Laboratory, Theoretical Division, Los Alamos, NM 87545, USA}

\vspace*{1cm}

\begin{abstract}
We present new results for the suppression of high transverse momentum charmonium 
[$J/\psi, \psi(2S)$] and bottomonium  [$\Upsilon(1S),\Upsilon(2S),\Upsilon(3S)$] states  in 
Pb+Pb  collisions at the Large Hadron Collider. Our theoretical formalism combines 
the collisional dissociation of quarkonia, as they propagate in the quark-gluon plasma, with the 
thermal wavefunction effects due to the screening of the $Q\bar{Q}$ attractive potential in  
the medium. We find that a good description of the relative suppression of the ground and
higher excited quarkonium states, transverse momentum and centrality distributions
is achieved, when comparison to measurements at
a  center-of-mass energy of 2.76~TeV is performed. Theoretical  predictions  for the highest  
Pb+Pb  center-of-mass energy  of 5.02~TeV at the LHC, where  new experimental results 
are being finalized, are  also presented.           
\end{abstract}


\maketitle
\section{Introduction~\label{section:Introduction}}

The fate of quarkonia -- for example the $J/\psi$ and the $\Upsilon$ meson families
-- in a thermal medium, such as the quark-gluon plasma (QGP) created in heavy
ion collisions (HIC), can help us characterize its properties. In particular, quarkonia are
sensitive to the space-time temperature profile and  transport coefficients of the QGP,
see~\cite{Mocsy:2013syh,Datta:2014wga,Andronic:2015wma} for
recent reviews. Experimentally, a key observable that carries such information is the nuclear
modification factor of the yields of quarkonia in nucleus-nucleus ($AA$) collisions, when
compared to their yields 
in nucleon-nucleon ($NN$) collisions scaled with the number of binary interactions
\begin{equation}
  R_{AA}= \frac{1}{\langle N_{\rm coll.} \rangle }
  \frac{d \sigma_{AA}^{\rm Quarkonia}/dy dp_T}{d \sigma_{pp}^{\rm Quarkonia}/dy dp_T}   
\label{raa}  
\end{equation}

In HICs one expects that the short distance formation dynamics of a $Q\bar{Q}$
pair is not affected
since $m_Q\gg T$ where $T$ is the typical temperature of the QGP. To simplify
the calculations, it is often also assumed that the matrix
elements for transition of $Q\bar{Q}$ to quarkonia is not modified, and
for every binary collision the formation of a specific quarkonium state
happens with the same probabilities as in $NN$ collisions. This takes a time scale close
to its inverse of its binding energy.  However, due to the screening of the
color interaction between $Q$ and $\bar{Q}$ in a deconfined
QGP~\cite{Matsui:1986dk}, as well as processes leading to the
dissociation~\cite{Xu:1995eb} of quarkonium states, we expect the yields of
quarkonia to be suppressed in heavy ion collisions ($R_{AA}<1$).

Several methods have been used to estimate screening and dissociation effects
encountered by a quarkonium in a thermal medium. A widely used approach is
based on the intuitive idea that the real part of the ``finite temperature
potential'' between two (nearly) static heavy quarks captures the screened
$Q\bar{Q}$ interaction while the imaginary part of the potential captures 
dissociation. For $T=0$, the real part can be quantitatively obtained  by calculating the 
Polyakov loop correlation functions ~\cite{Kaczmarek:2005ui,Bazavov:2012bq,Burnier:2015tda}.
For $T>0$, the connection between various correlators
 calculated on the lattice
and the potential between $Q$ and $\bar{Q}$ is
subtle~\cite{Kaczmarek:2005ui,Mocsy:2007jz,Bazavov:2012bq,Burnier:2015tda}.
While the singlet free energy $F_1(r)$ of the $Q\bar{Q}$ state as a function of
the separation $r$, and internal energy $U(r)$ can be extracted from the
lattice data, the connection of either of the two with the $Q\bar{Q}$ potential
is indirect. An important step in clarifying this connection was taken in
Ref.~\cite{Laine:2006ns}, which showed that
the ``potential'' between two heavy quarks is complex, with the imaginary part
connected to thermal processes that can lead to dissociation of quarkonia.
Significant progress has been made in the perturbative calculation of the real and
imaginary part of the $Q\bar{Q}$ potential for $Q\bar{Q}$ at rest in the 
medium~\cite{Laine:2006ns,Margotta:2011ta,Brambilla:2013dpa} 
or moving slowly in the medium~\cite{Escobedo:2013tca}. Analytic calculations can be performed in
certain regimes by considering various hierarchies of energy scales. For example, one has to assume
that $T$ and $m_Q$ are large enough so that perturbation theory is valid all the way down to
energy scales $\pi T$ and the binding energy $E_b$.

While the real part of the potential for $T\gg \Lambda_{QCD}$ can be obtained
using perturbation theory, 
non-perturbative effects are substantial near the crossover temperature
and it is better to estimate this quantity
from lattice calculations by matching the Euclidean correlators measured
on the
lattice~\cite{Mocsy:2007jz,Rothkopf:2009pk,Petreczky:2010tk,Burnier:2015tda}
with those evaluated using the potential. The extraction of the
imaginary part of the potential using this technique is challenging and often
perturbative estimates are used. Extensive
phenomenological study of quarkonium suppression by using this approach has
been performed~\cite{Strickland:2011mw,Song:2014qoa,Krouppa:2015yoa,Hoelck:2016tqf}.
Recently, approaches treating the $Q\bar{Q}$ as an open quantum system were
developed, where a stochastic equation is written for the evolution of the $Q\bar{Q}$
wavefunction~\cite{Akamatsu:2011se,Blaizot:2015hya,Brambilla:2016wgg,Kajimoto:2017rel}.
We also note that the connection between the
heavy quark correlators measured
on the lattice and in a non-perturbative framework using the
in medium T-matrix has been made in Refs.~\cite{Riek:2010fk,Riek:2010py}. 
Quarkonium suppression~\cite{Rapp:2009my,Emerick:2011xu,Du:2017qkv}, as well as  
low $p_T$ observables like heavy quark flow have been studied using this approach. 
Thermal properties of quarkonia have also been investigated in the strong
coupling regime using AdS/CFT techniques, see Ref.~\cite{CasalderreySolana:2011us}.

In this paper we calculate the differential $R_{AA}$ as a function of $p_T$ by
solving rate equations~\cite{Adil:2006ra,Sharma:2009hn,Sharma:2012dy}
describing the change in the yields as a function of time in HICs.
Conceptually, our approach resembles treating the $Q\bar{Q}$ as an open quantum
system. It was, however,  introduced earlier to describe the attenuation of open heavy
flavor~\cite{Adil:2006ra}. Gluon exchanges with the medium lead to the modification of the
$Q\bar{Q}$ state and, hence, a reduction in the overlap with the quarkonium
wavefunction. However we do not connect the imaginary part of the potential
calculated in approaches cited above to the decay rate. This is because we
focus on $R_{AA}$ at high $p_T$. For high $p_T$ partons traversing the medium,
a very successful picture is that the interactions with the medium lead to
transverse momentum broadening. The decay rate in our calculation is related to
the accumulation of relative momenta between $Q$ and $\bar{Q}$.

The form of the rate equations is the same as used in our previous
work~\cite{Sharma:2012dy} and they involve the dissociation time and the formation
time as inputs. The formation time in our formalism is a measure of the time
scale on which the proto-quarkonium $Q\bar{Q}$ state develops interactions
with the medium  and its overlap  with the
quarkonia becomes substantial. We vary it in the neighborhood of ${\calO}(1\,
{\rm{fm}})$. The dissociation time scale is computed as in~\cite{Sharma:2012dy}
by calculating the survival probability of the quarkonia.

One conceptual change in our framework, when compared to~\cite{Sharma:2012dy},
is that we use the real part of the lattice motivated thermal
potentials~\cite{Mocsy:2007jz,Petreczky:2010tk,Bazavov:2012bq} to solve for the
quarkonium wavefunction and square the overlap with the thermal wavefunction to
get the survival probability. This is justified if the time scales on which the
medium screens the $Q\bar{Q}$ interaction is smaller than the formation time as
well as the dissociation time. A rough estimate of the Debye screening time is
the inverse of the Debye screening mass ($\mu_D$), which is $\sim gT$. For
$g=1.85$, ($1/\mu_D$) is numerically smaller than $1/2$~fm for most of the
evolution of the medium at the LHC.  On the other hand, the formation is taken
to be $\sim 1$~fm. Finally, this puts a constraint that the minimum
dissociation time consistent within our formalism is roughly $1$~fm. We see
that while there is no substantial hierarchy between $1/\mu_D$ and the
formation and dissociation time scales in our calculation, numerically
$1/\mu_D$ is smaller and, hence, we work in this approximation. 

In addition, we make two technical improvements. First, we consider a $2+1$ dimensional
viscous hydrodynamic medium, rather than a simplified $1+1$ dimensional Bjorken
expansion, as a model for the QGP~\cite{Shen:2014vra}.  An important element of
our approach is the use of non-relativistic QCD (NRQCD)~\cite{Bodwin:1994jh} to
obtain the  baseline nucleon-nucleon cross sections for quarkonia  and understand the
$p_T$-dependent feed down. The second technical improvement is a refit of the
long distance matrix elements for $\chi_c$ and $\psi(2S)$ to obtain a
better description of the high
$p_T$ part of their differential spectra. We are further
motivated by the  fact that new experimental results that extend the
measurements of quarkonia to high $p_T$ have appeared. Improved measurements
of the  higher excited states, such as the $\psi(2S)$ and the $\Upsilon(2S)$
and $\Upsilon(3S)$, that are differential in transverse momentum are
particularly useful in constraining the mechanisms of quarkonium suppression.

In this paper we focus on the nuclear modification of the prompt $J/\psi$ and $\Upsilon$
states around mid-rapidity at the LHC and at high transverse
momentum~\cite{Adam:2015rba,Abelev:2013ila,Aad:2010aa,Khachatryan:2016ypw,Khachatryan:2016xxp,Sirunyan:2016znt}. In section~\ref{section:formalism} we describe the theoretical model that is
employed in the calculation of quarkonium production in  heavy ion
collisions. Phenomenological results for the nuclear modification of the
$J/\psi$ and $\Upsilon$ states are presented in section~\ref{section:Results}.
Our conclusions and outlook are given in section~\ref{section:Conclusions}.

\section{Theoretical formalism~\label{section:formalism}}

Quarkonia are bound states of heavy quarks ($Q$) and antiquarks ($\bar{Q}$),
and a concrete picture of the dynamics of the heavy quark pair ($Q\barQ$) in
$NN$ collisions is given by non-relativistic quantum chromodynamics
(NRQCD)~\cite{Bodwin:1994jh}. In this effective theory, the initial hard
collision produces a short distance ($\sim1/m_Q$) $Q\barQ$ pair in a
color-singlet or an octet state with a specific spin and orbital structure. The
production cross-section for this short-distance state can be calculated using
perturbative QCD. This $Q\bar{Q}$ state evolves into a quarkonium state with
probabilities that are given by long distance matrix elements (LDMEs). For
color-octet states, this evolution process also involves the emission of soft
partons to form a net color-singlet object which we assume occurs on a
time scale which is shorter than ${\cal O}(1\, {\rm fm})$.

This framework has been successfully used to calculate the
unpolarized differential yields of quarkonia versus the transverse momentum
($p_T$) in proton-proton ($pp$) or proton-antiproton ($p\bar{p}$)]
collisions~\cite{Cho:1995vh,Braaten:2000cm,Butenschoen:2011yh,Wang:2012is}.
An accurate description of both the cross sections and  polarization of
quarkonia in hadronic reactions still remains a challenge~\cite{Bodwin:2014gia}.
Recently, it has been suggested~\cite{Baumgart:2014upa,Kang:2017yde,Bain:2017wvk}
that new experimental measurements of quarkonium production inside
jets~\cite{Aaij:2017fak} may help  better constrain the relevant LDMEs.  The
focus of  this paper is, however, different.  Our study concentrates on the
production, propagation, and dissociation of quarkonium states in strongly
interacting matter.  We follow the NRQCD calculation outlined
in~\cite{Sharma:2012dy} and use the LDMEs extracted there to give good
description of the cross sections for bottomonia for $pp$ and
$p\bar{p}$ collisions for $p_T$ in the range of $5$ to $30$~GeV. For charmonia
we improve the $\chi_c$ fitting procedure by allowing the singlet matrix
element as a free parameter. We also refit LDMEs for the $\psi(2S)$ by fitting
the LHC 7~TeV and CDF 1.8~TeV data. Both these changes make the spectra for
$\chi_c$ and $\psi(2S)$ in the $p_T\sim 10-20$~GeV region softer and improve the
description of data without spoiling the agreement at lower
$p_T$. 



In order to address quarkonium attenuation in heavy ion reactions, we need to 
understand the $J/\psi$ and $\Upsilon$ states' behavior and melting at finite
temperature in the QGP, and the dissociation processes 
due to collisional interactions with the quasi-particles of the QCD medium. 
 To accomplish this,  a detailed knowledge of the wavefunctions
 at zero and finite temperatures is necessary. We start by solving  the Schr\"{o}dinger
 equation  by  separating  the radial and angular parts of the wavefunction,
 $\psi(\bfr)=Y_l^m(\hatr)  R_{nl}(r)$. 
The reduced equation for the  radial part  can be written as
\begin{eqnarray}
&& \left[-\frac{1}{2\mu_{\rm{red}}}\frac{\partial^2}{\partial r^2}
 +\frac{l(l+1)}{2\mu_{\rm{red}} r^2}
 +V(r)\right]rR_{nl}(r)  \nonumber \\[1ex]
 && = (E_{nl}) rR_{nl}(r)  \; , 
\end{eqnarray}
where $\mu_{\rm{red}}=\frac{m_Q}{2}$ is the reduced mass, and $n$ and $l$
are the principal and orbital quantum numbers, respectively.  
$V(r)$ is the potential between the two heavy quarks, which can be estimated
from the lattice~\cite{Bazavov:2012bq}.  The binding energy of the meson is
then $E^b_{nl}=V(\infty) - E_{nl}$.  For $T=0$ we take the form of the potential to
be of the standard Cornell form 
\begin{equation}
\begin{split}
V(r) = - \frac{\xi}{r} + \sigma r , \;\; \xi=0.385,  \;  \sigma=0.224\;{\rm{GeV}}^2.
\end{split}
\end{equation}
The form of the potential has been validated in multiple lattice calculations 
by evaluating the Polyakov loop correlator as a function of the quark-antiquark
separation. The Cornell long-distance  part is cut off at
$r_{\rm{max}}=1.1$~GeV~\cite{Mocsy:2007jz} to
model string breaking on this length scale. This defines the value of
$V(\infty)\approx 1.2$~GeV. For the mass of the heavy quark we take 
$m_c=1.34$ for the charm quark and $m_b=4.5$ for the bottom quark.
For $T>0$ the form of the real part of the  potential we use is the internal
energy found in~\cite{Kaczmarek:2005ui}. 
We have checked that the potentials we use are quite close to the internal energies
computed in the more recent work~\cite{Bazavov:2012bq} and the potentials used
in~\cite{Petreczky:2010tk}. Internal energies provide a stronger binding potential
compared to the single
free energies~\cite{Bazavov:2012bq}. The use of the internal energies can be
justified if the dynamics of the quarkonia 
are too rapid for the $Q\bar{Q}$ potential to fully thermalize, which might be
especially likely for high $p_T$ quarkonia in the medium.  With this setup,
solutions are obtained for a variety of temperatures and for the  S-wave and
P-wave states.  Even though we are primarily interested in the $J/\psi$,
$\psi(2S)$, and $\Upsilon(nS) $, the   P-wave $\chi_c$ and $\chi_b$ contribute
via feed down. 

\begin{table}[h]
\begin{tabular}{cccccc}
$l$ & $n$ & $E^b_{nl}$~(GeV) & $\sqrt{\langle r^2\rangle}$~(GeV$^{-1}$)  & $k^2$~(GeV$^2$) & Meson \\ 
\hline \hline
0 & 1 & 0.700 & 2.24 & 0.30 & $J/\psi$ \\
0 & 2 & 0.086 & 5.39 & 0.05 & $\psi(2S)$ \\
1 & 1 & 0.268 & 3.50 & 0.20  & $\chi_{c}$ \\[1ex]
\hline
0 & 1 & 1.122  & 1.23 & 0.99 & $\Upsilon(1S)$ \\
0 & 2 & 0.578  & 2.60 & 0.22 & $\Upsilon(2S)$ \\
0 & 3 & 0.214  & 3.89 & 0.10 & $\Upsilon(3S)$ \\
1 & 1 & 0.710  & 2.07 & 0.58 & $\chi_{b}(1P)$ \\
1 & 2 & 0.325  & 3.31 & 0.23 & $\chi_{b}(2P)$ \\
1 & 3 & 0.051  & 5.57 & 0.08 & $\chi_{b}(3P)$\\
\end{tabular}
\caption{Charmonia and bottomonia wavefunctions at zero temperature. $l$ refers to the angular momentum of the
$Q\barQ$ state, while $n$ is the radial quantum number. $E^b_{nl}$ is the binding energy, $\sqrt{\langle r^2\rangle}$ is the root mean square  (RMS)
radius of the quarkonium state, and  $k^2$ is the mean square momentum. }\label{table:Tzero}
\end{table}

\begin{table}[h]
\begin{tabular}{cccccc}
$l$ & $n$ & $E^b_{nl}$~(GeV) & $\sqrt{\langle r^2\rangle}$~(GeV$^{-1}$)  & $k^2$~(GeV$^2$) & Meson \\ 
\hline \hline
0 & 1 & 0.366 & 2.34  & 0.27  & $J/\psi$  \\
0 & 2 & - & -  & -  & $\psi(2S)$ \\
1 & 1 & 0.003 & 8.15  & 0.04  & $\chi_{c}$  \\[1ex]
\hline
0 & 1 & 0.782  & 1.23  &  0.98  & $\Upsilon(1S)$ \\
0 & 2 & 0.244  & 2.72  &  0.20       & $\Upsilon(2S)$ \\
0 & 3 & -      & -     &  -       & $\Upsilon(3S)$ \\
1 & 1 & 0.371  & 2.09  &  0.57        & $\chi_{b}(1P)$ \\
1 & 2 & 0.040  & 4.56  &  0.12        & $\chi_{b}(2P)$ \\
1 & 3 & -      & -     &  -       & $\chi_{b}(3P)$ \\
\end{tabular}
\caption{Charmonia and bottomonia  wavefunctions at finite temperature. We chose a temperature  of 192~MeV  to 
illustrate the  disappearance of the weakly bound states and the changes in the quarkonium wavefunctions.}
\label{table:Tfinite}
\end{table}

Pertinent results  for the  quarkonium and bottomonium wavefunctions are
presented in Table~\ref{table:Tzero} and Table~\ref{table:Tfinite} for zero
temperature ($T$) and $T=192$~MeV, respectively. We show the binding energy
$E^b_{nl}$ and  the root mean square  (RMS) size $\sqrt{\langle r^2\rangle}$ of
the quarkonium state. The RMS size changes only slowly with $T$, except near
the  dissociation temperature. We have chosen the finite temperature value
$\sim 190$~MeV to illustrate that several states, i.e.  $\psi(2S)$,
$\Upsilon(3S)$, and  $\chi_{b}(3P)$, cease to exist even with our selected  strong
binding potential. The $\chi_{c}$  and $\chi_{b}(2P)$ states are very near
dissociation.  If we Fourier transform to momentum space, the mean squared
momentum  $k^2$ can also be evaluated and is given in  Tables~\ref{table:Tzero}
and~\ref{table:Tfinite}. We note that while there is a correlation between the 
width of the wavefunctions  of quarkonia and their binding energies, this 
correlation is highly non-linear. The widths change rapidly only near 
dissociation when $E^b_{nl} \rightarrow 0$.

In this work we are interested in large transverse momentum quarkonia and it is convenient to work in light
cone momenta and with light-cone wavefunctions. The relation between the instant form and the light-cone form  
of the momentum space wavefunctions for mesons was discussed in detail  in~\cite{Sharma:2009hn,Sharma:2012dy}. 
For the case of quarkonia,  the color-singlet contribution  can be understood  as one  matching to the lowest order ($n=2$) 
Fock component of the state.  A color-octet initial state  must emit  at least one gluon for a color neutral hadron to be
produced. In either case the heavy meson or proto-quarkonium  state of momentum $\vec{P^+} = (P^+,{\bf P})$ can be 
approximated as:
\begin{eqnarray}       
 |\vec{P}^+ \rangle 
&=& \int \frac{d^2{\bf k}}{(2\pi)^{3}} \frac{dx}{ 2\sqrt{ x(1-x)}}
  \frac{\delta_{c_1c_2}}{\sqrt{3}} \, 
\psi(x,{\bf k})  \nonumber \\
&& \times a_Q^{\dagger\;  c_1 }(x\vec{P}^++{\bf k})  b_{\bar{Q}}^{\dagger \;  c_2  }
((1-x)\vec{P}^+-{\bf k})  |0 \rangle  \; , 
\label{Mp1}
\end{eqnarray}       
where  $a^\dagger$ ($b^\dagger$) represent an 
``effective'' heavy quark (anti-quark) in the $3$ ($\bar{3}$) state, $c_1,c_2$ being the color 
indices~\cite{Sharma:2009hn,Sharma:2012dy}. 
The   light cone wavefunction $\psi(x, {\bf{k}})$ in Eq.~(\ref{Mp1}), which describe the longitudinal momentum 
fraction $x$ ($1-x$) distribution  and the transverse momentum ${\bf k}$ ($-{\bf k}$) distribution of 
heavy quarks (antiquarks) 
is given by  
\begin{eqnarray}
  && \psi(x,{\bf k}) = {\rm Norm} \times \exp\left( - \frac{{\bf k}^2 + m_Q^2 }{2 \Lambda^2(T) x(1-x) }   
  \right) \;   ,  \nonumber \\
&& \frac{1}{2 (2\pi)^{3} } \int dx d^2{\bf k}  \;
| \psi(x,{\bf k}) |^2 = 1 \; .
\label{lonorm}
\end{eqnarray}
In Eq.~(\ref{lonorm}) $\Lambda(T)$ is the transverse momentum width of the light-cone wavefunction which
needs to be constrained at 0 and finite temperature to the mean transverse momentum squared
from the solution to the Schr\"{o}dinger equation, which remains invariant under boost.   
If we introduce the notation $\Delta {\bf k} = {\bf k}_1-{\bf k}_2 = 2 {\bf
k}$, the equation for  $\Lambda(T)$ reads
\begin{equation}
\frac{1}{2 (2\pi)^{3} } \int dx d^2{\bf k}  \;
{\Delta \bf k}^2  | \psi(x,{\bf k}) |^2 =  4\langle {\bf k^2} \rangle =  \frac{2}{3} \kappa^2  \; .
\label{lonorm2}
\end{equation}
The factor $2/3$ comes from the 2D projection of the mean squared  transverse
momentum $\kappa^2$ from the instant-form wavefunction.

\begin{figure}[!t]
\vspace*{-.1in}
\hspace*{-.15in}\includegraphics[width=3.5in,angle=0]{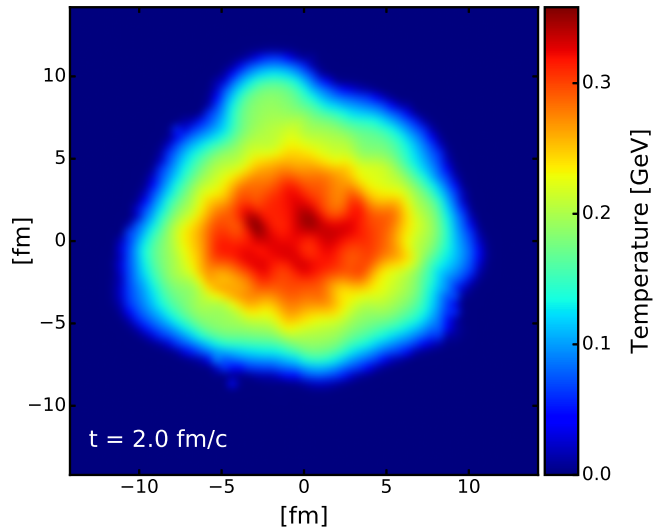} 
\caption{(Color online) Example of the temperature profile of the QGP at a typical time $t=2$~fm
on 0-10\% central Pb+Pb collisions at $\sqrt{s} = 2.76$~TeV the LHC. Glauber initial conditions and $\eta/s=0.08$ are used
in a 2+1D hydrodynamic simulation based on~\cite{Shen:2014vra}. }
\label{hydro}
\end{figure}

The temperature and/or density profiles of the medium, which play an important role in the
dissociation of quarkonia, can be obtained from hydrodynamic simulations of the 
QGP~\cite{Schenke:2010nt,Habich:2014jna,Shen:2014vra} in $2.76$~TeV and $5.02$~TeV
Pb+Pb collisions at the LHC. Specifically, we use the iEBE-VISHNU (2+1)-dimensional 
event-by-event viscous hydrodynamic package~\cite{Shen:2014vra}. A sample
temperature distribution at time $t=2$~fm when the interplay between the formation and dissociation of quarkonia  
is important in setting the final observed $J/\psi$ and $\Upsilon$ multiplicities  is shown in 
Fig.~\ref{hydro}.  By comparing the temperature in the different points in the $(x,y)$ plane
perpendicular to the collision axis to the results in Table~\ref{table:Tfinite} one can get a sense of
how the different quarkonium states  will be attenuated in heavy ion collisions relative to proton collisions.

The propagation of a  $Q\bar{Q}$ state in matter is accompanied by collisional
interactions mediated at the partonic level, as long as the momentum exchanges between the
medium quasi-particles and the heavy quarks can resolve the partonic structure of the  meson.  The related 
modification of the quarkonium wavefunction in Eq.~(\ref{lonorm})  can lead to the dissociation of  
$J/\psi$s and $\Upsilon$s in addition to the thermal 
effects. The cumulative one dimensional momentum transfer for a quarkonium state that 
starts at transverse position ${\bf x}_0$ and propagates with velocity ${\bm \beta}$,
such that  ${\bf x}(\tau) = {\bf x}_0 + {\bm \beta} (\tau-t_0)$,  reads
\begin{equation}
\chi \mu_D^2 \xi  = \int_{t_0}^t  d\tau 
\frac{\mu_D^2(\bf{x}(\tau),\tau) }{\lambda_q({\bf x}(\tau),\tau) } \xi\, .
~\label{eq:xibar}
\end{equation}
Here $\mu_D^2 = g^2T^2(1+N_f/6)$   is the  Debye screening scale and we use 2 active light quark flavors
$N_f=2$.  The scattering inverse length of the quark is ${1}/{\lambda_q}  = \sigma_{qq}\rho_q + \sigma_{qg} \rho_g$,
where $\rho_q$ and $\rho_g$ are the partial densities of light quarks and gluons in the QGP.   We label the  
cumulative one dimensional momentum transfer $\chi \mu_D^2 \xi$ in analogy with a uniform static medium
 where the Debye scale is fixed and $\chi = L/\lambda_q$ is the opacity. The elastic scattering cross sections are given by 
\begin{equation}
\sigma_{qq} = \frac{1}{18 \pi} \frac{g^4}{\mu_D^2} \, ,  \quad \sigma_{qg} = \frac{1}{8 \pi} \frac{g^4}{\mu_D^2} \, .
\end{equation}
Last but not least, $\xi$ is a parameter related to the heavy quark broadening from multiple scattering in the 
QGP~\cite{Gyulassy:2002yv,Adil:2006ra}.  In the limit of strictly soft interactions $\xi=1$ and an enhancement
of $\xi  \sim$few  may arise from the power law tails of the Moliere
multiple scattering. 

\begin{figure}[!t]
\vspace*{-.0in}
\hspace*{-.15in}\includegraphics[width=3.4in,angle=0]{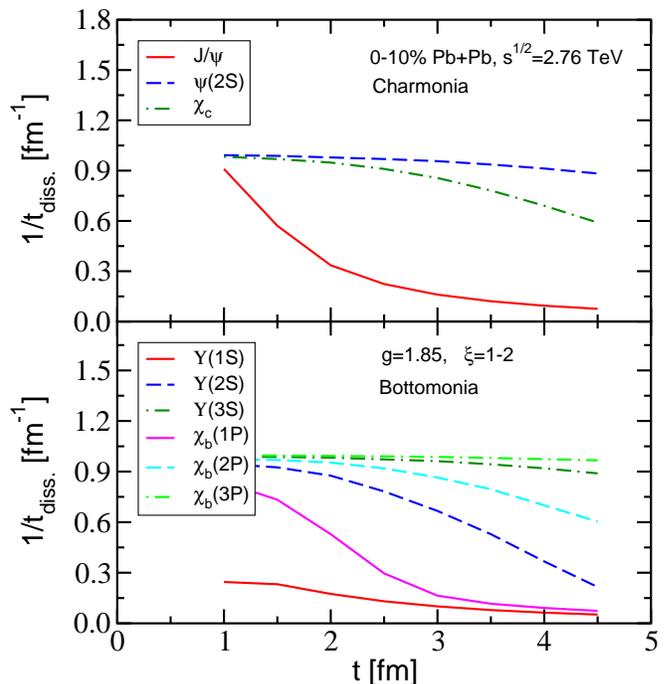} 
\caption{(Color online)  The dissociation rate $1/\tau_{\rm diss.}$  for charmonium and bottomonium states
in 0-10\% central Pb+Pb collisions at $\sqrt{s}=2.76$~TeV is shown in the top and bottom panels, respectively.
We select the coupling between the quarks and the medium $g=1.85$ and the broadening parameter $\xi=1$.
}
\label{rate}
\end{figure}

We initialize the wavefunction $\psi_{i}(\Delta {\bf k}, x)$ of the proto-quarkonium $Q\bar{Q}$  state with a 
width $\Lambda_0 \equiv \Lambda(T=0) $. 
This is a natural choice since in the absence of a medium it will evolve on the time-scale of
${\cal O}(1{\rm fm})$ or greater into the 
observed heavy meson. By propagating in the medium this initial wavefunction  accumulates transverse momentum 
broadening  $\chi \mu_D^2 \xi$.  The probability that this  $Q\bar{Q}$ configuration will transition into a
final-state heavy meson 
with thermal wavefunction $\psi_{f}(\Delta {\bf k}, x)$ with $\Lambda(T)$ is given by
\begin{eqnarray}
  P_{f\leftarrow i} (\chi\mu_D^2 \xi,T) & = & \left|  \frac{1}{2 (2\pi)^{3} }  \int d^{2}{\bf k} dx \,
\psi_{f}^* (\Delta {\bf k},x)\psi_{i}(\Delta {\bf k}, x) \right|^{2}  
\nonumber \\
&& \hspace*{-1.in}= \left| \frac{1}{2 (2\pi)^{3} }  \int dx \; {\rm Norm}_f {\rm Norm}_i \, \pi  
  \,  e^{-\frac{ m_{Q}^{2} }{ x(1-x)\Lambda(T)^{2} } }  e^{-\frac{ m_{Q}^{2} }{ x(1-x)\Lambda_0^{2} } }  \right.   \nonumber \\
&&  \hspace*{-.9in}  \times \left .  \,   \frac{ 2 [ x(1-x)\Lambda(T)^{2}]
[\chi\mu_D^{2}\xi+x(1-x)\Lambda_0^{2}] }
{   [ x(1-x)\Lambda(T)^{2}] + [\chi\mu_D^{2}\xi+x(1-x)\Lambda_0^{2}]  }  \; \right|^2 \, . \;\; \quad 
\label{sprob}
\end{eqnarray}
In Eq.~(\ref{sprob}) ${\rm Norm}_i$  is the normalization of the initial state,
including the transverse momentum broadening 
from collisional interactions,  and  ${\rm Norm}_f$ 
is the normalization of the final state. 
The dissociation rate for the specific quarkonium state  can then be expressed as 
\begin{equation}
\frac{1}{t_{\rm diss.}}  =  -  \frac{1}{P_{f\leftarrow i} (\chi\mu_D^2 \xi,T)}  
 \frac{d P_{f\leftarrow i} (\chi\mu_D^2 \xi,T) }{dt}  \, .
\end{equation}
It will enter the time evolution of the $J/\psi$s and $\Upsilon$s in the medium. 
To visualize the dissociation rate,  we  present  $1/\tau_{\rm diss.}$ as a function of time in Fig.~\ref{rate} 
for 0-10\% central Pb+Pb collisions at the LHC at $\sqrt{s}$  per nucleon pair of 2.76 TeV. 
We have weighed the dissociation rates with the binary collision density in the plane perpendicular to the
collision axis and averaged over multiple hydrodynamic events. The top panel shows the dissociation rates of 
charmonium states and the bottom panel shows the corresponding rates for bottomonium state.  
We find that the modification of the ground states, such as $J/\psi$, $\Upsilon(1S)$ is dominated 
by dissociation at the early stages of the evolution of the QGP,  whereas excited weakly bound states, such as  are  
$\psi(2S)$, $\chi_b(3P)$, can be strongly modified until they escape the medium.

Finally, we note that effects that are suppressed at high transverse momentum,
such as recombination~\cite{Thews:2006ia} of unbound $Q$ and $\bar{Q}$ because
of the Boltzmann suppression factor~\cite{Gupta:2014ova}, the Cronin
effect~\cite{Qiu:2003pm}, and power corrections~\cite{Qiu:2004qk} do not play a
role. Cold nuclear matter (CNM) energy loss might affect production cross
sections~\cite{Vitev:2006bi},  but its effects become significant at very high
$p_T$ near the kinematic threshold~\cite{Kang:2015mta}.  Last but not least, it
was also found that leading-twist shadowing effects near mid-rapidity at high
transverse momentum at the LHC are small~\cite{Vogt:2015uba}.  While
experimental results in $p+$Pb collisions at the LHC cannot exclude CNM effects
at the $5-10\%$ level at $p_T>5$~GeV around mid-rapidity, they are also consistent
with no nuclear
modification~\cite{Adam:2015iga,Sirunyan:2017mzd,Aaboud:2017cif}.  For
these reasons, for our study at $|y|<2.4$, $p_T > 5$~GeV we neglect these
effects.

\section{Phenomenological results~\label{section:Results}}

\begin{figure}[!t]
\vspace*{.0in}
\includegraphics[width=3.38in,height=3.5in]{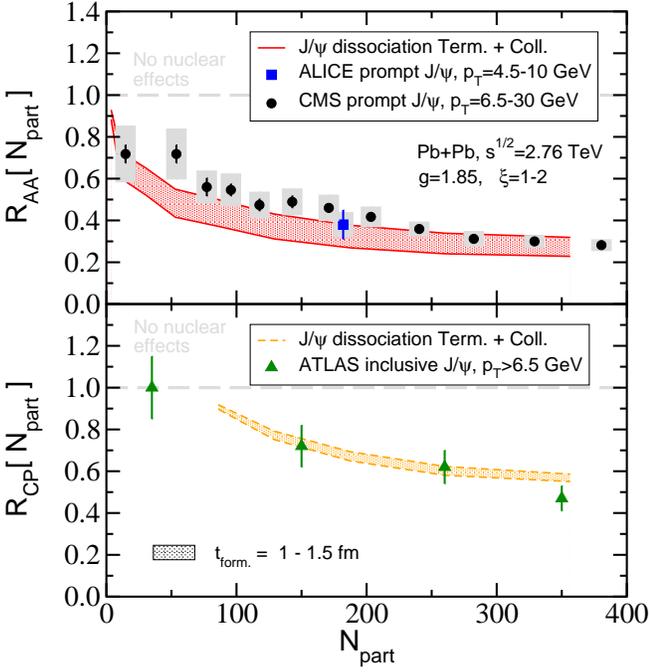} 
\caption{(Color online) Comparison of theoretical results for prompt
  $J/\psi$ suppression for $p_T> 6.5$~GeV to LHC Pb+Pb results at
  $\sqrt{s}=2.76$~TeV. Top panel: $N_{\rm part.}$ dependence versus
  ALICE~\cite{Adam:2015rba} and CMS measurements~\cite{Khachatryan:2016ypw}.
  Bottom panel: $R_{CP}$ as a function of $N_{part}$ versus ATLAS
  measurements~\cite{Aad:2010aa}.}
\label{276Npart}
\end{figure}

In this section we present the phenomenological results of our theoretical
model for quarkonium  dissociation due to thermal wavefunction effects and
collisional breakup.  For every centrality class and hydrodynamic event we
distribute the production of the proto-quarkonium states according to the
binary collision density in the 2D plane perpendicular to the collision axis.
The azimuthal distribution  in the directions of quarkonium propagation is
uniform and we also average over multiple  fluctuating hydrodynamic events that
describe the QGP background.

For each quarkonium state the dynamics of production and propagation through the QCD
medium described above is given by a set of differential equations: 
\begin{eqnarray}
\label{rateq11}
\frac{d}{dt} \left( \frac{d\sigma^{Q\bar{Q}}(t;p_T)}{dp_T} \right) 
&=& - \frac{1}{t_{\rm form.} } \frac{ d\sigma^{Q\bar{Q}}(p_T)}{dp_T}  \, , \\
\frac{d}{dt} \left(  \frac{d\sigma^{\rm meson}(t;p_T)}{dp_T} \right) 
\label{rateq12}
&=&  \frac{1}{t_{\rm form.}} \frac{ d\sigma^{Q\bar{Q}}(t;p_T)}{dp_T}  \nonumber \\
&&- \frac{1}{t_{\rm diss.}}  \frac{d\sigma^{\rm meson}(t;p_T)}{dp_T}   \, , \\
\frac{d}{dt} \left(  \frac{d \sigma^{\rm diss.} (t;p_T)}{dp_T} \right) 
&=&  \frac{1}{t_{\rm diss.}}   \frac{d\sigma^{\rm meson}(t;p_T)}{dp_T}  \, . 
\label{rateq13}
\end{eqnarray}
Here we denote by ${d\sigma^{Q\bar{Q}}(t;p_T)}/{dp_T}$ the cross section to produce the 
proto-quarkonium states that evolve into an interacting with the medium meson on the time scale of $t_{\rm form.}$.
This time scale for the heavy quarks to interact with the QGP is taken in the relativistic limit.   
The initial condition at $t\approx 0$ includes the short distance perturbative  $Q\bar{Q}$ production cross sections and the 
long-distance matrix elements for the particular quarkonium state.   ${d\sigma^{\rm meson}(t;p_T)}{dp_T}$ is the
cross section for that quarkonium state as a function of time $t$. Finally,  ${d \sigma^{\rm diss.} (t;p_T)}/{dp_T}$ is the cross section of the dissociated 
$Q\bar{Q}$ pairs that will not produce quarkonia. In the absence of a medium $t_{\rm diss.} \rightarrow \infty$ and the proto-quarkonium 
states evolve into the corresponding meson with unit probability.  The system of equations Eqs.~(\ref{rateq12})-(\ref{rateq13})  
has to be solved for each of the quarkonium states  listed in Table~\ref{table:Tzero}.
Feed-down is then performed using the NRQCD cross sections and branching ratios.

\begin{figure}[!t]
\vspace*{.0in}
\includegraphics[width=3.38in,height=2.2in]{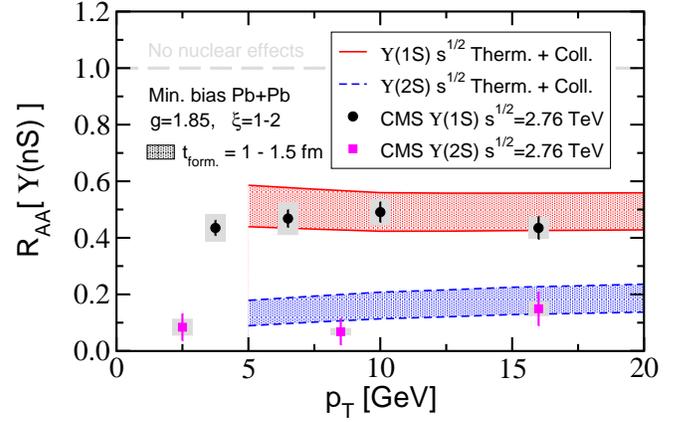} 
\caption{(Color online) Comparison of theoretical results for the
  $\Upsilon(nS)$ $R_{AA}$ in 2.76~TeV minimum bias Pb+Pb collisions versus $p_T$
to CMS experimental measurements~\cite{Khachatryan:2016xxp}.
}
\label{U276}
\end{figure}

\begin{figure}[!t]
\vspace*{.0in}
\includegraphics[width=3.38in,height=2.2in]{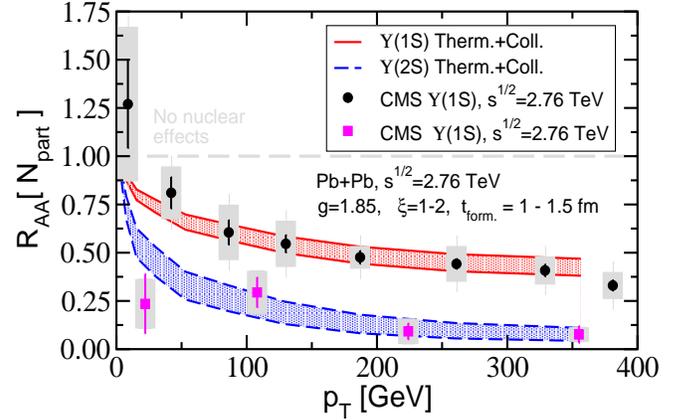} 
\caption{(Color online) Suppression of $\Upsilon(1S)$, $\Upsilon(2S)$ in Pb+Pb collisions
at  2.76~TeV is shown versus the number of participants for  $p_T=5$~GeV.
Superimposed are  CMS experimental  data~\cite{Khachatryan:2016xxp} on bottomonia suppression
versus centrality. 
}
\label{U276cent}
\end{figure}

We start by first discussing results at the lower energy of $\sqrt{s}=2.76$~TeV. 
In Fig.~\ref{276Npart} we present the centrality dependence of prompt $J/\psi$s
in Pb+Pb collisions. The bands reflect the combined uncertainty
of the interaction onset time $t_{\rm form.}$ and the collisional dissociation of
the quarkonium states. In the evaluation of the latter we keep the coupling between
the heavy quarks and the medium fixed at $g=1.85$~\cite{Sharma:2012dy} but vary the
broadening parameter $\xi$. The upper edge of the uncertainty band corresponds to
$t_{\rm form.} = 1.5$~fm, $\xi =1$. The lower edge of the uncertainty band corresponds
to $t_{\rm form.} = 1$~fm, $\xi =2$. The upper panel of  Fig.~\ref{276Npart} shows
comparison to the ALICE~\cite{Adam:2015rba} and CMS~\cite{Khachatryan:2016ypw} prompt
$J/\psi$ measurements. We find improved description in the most central $N_{\rm part.}$
bins relative to the case when thermal screening effects were not
included~\cite{Sharma:2012dy}. Around $N_{\rm part.} =100$ the data lies on the
upper edge of the theoretical error band. The bottom panel of Fig.~\ref{276Npart} shows the $J/\psi$
$R_{CP}$, where the 40\%-80\% peripheral collisions are used as a baseline.
The  ATLAS collaboration measured inclusive $J/\psi$~\cite{Aad:2010aa}. However, in the
$p_T < 10$~GeV interval which dominates the cross section, the non-prompt
$B\rightarrow J/\psi$ contribution is limited to 20-30\%~\cite{Sharma:2012dy} and will
not noticeably  affect the theoretical results.

\begin{figure}[!t]
\vspace*{.0in}
\includegraphics[width=3.38in,height=3.5in]{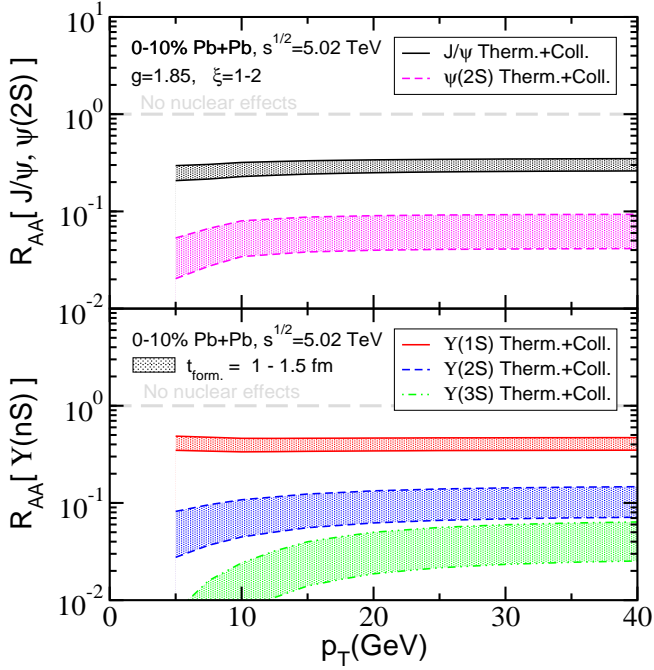} 
\caption{(Color online) Theoretical model predictions for the $R_{AA}$ of the ground
  and excited $J/\psi$ (top panel) and $\Upsilon$ (bottom panel) states
  in 0-10\% central  Pb+Pb collisions  at  $\sqrt{S}=5.02$~TeV at the LHC. The
  coupling between the heavy quarks and the medium $g=1.85$ and the bands correspond to
$t_{\rm form.}=1.5$~fm, $\xi=1$ - $t_{\rm form.}=1$~fm, $\xi=2$. }
\label{502cen}
\end{figure}

Recently, experimental results for the differential  suppression of the $\Upsilon(nS)$
family have appeared at high $p_T$~\cite{Khachatryan:2016xxp}. Theoretical calculations
for the $\Upsilon(1S)$ (red band) and $\Upsilon(2S)$ (blue band) in minimum bias
$\sqrt{s}=2.76$~TeV Pb+Pb reactions are shown in Fig.~\ref{U276}. We have evaluated
the cross sections for quarkonia in 10 centrality classes (labeled $i$) and 
\begin{eqnarray}
  R_{AA}^{\rm min.\ bias}(p_T)& =&
  \frac{\sum_i R_{AA}(\langle b_i \rangle) W_i}{\sum_i W_i} \quad {\rm where}
  \nonumber \\
  W_i &=&\int_{b_{i\, \min}}^{b_{i\, \max}}  N_{\rm coll.}(b)\,\pi \, b \,db  \, .
\label{cent}
\end{eqnarray}
The experimental data is described well, including its magnitude and $p_T$ dependence.
We note that collisional dissociation mostly        
affects the ground  $\Upsilon$ state, while thermal wavefunction effects dominate
the attenuation pattern of the excited $\Upsilon$ states. The CMS collaboration also
put an upper limit on the  $\Upsilon(3S)$ cross section in Pb+Pb reactions, corresponding to an
upper limit on its $R_{AA}$~\cite{Khachatryan:2016xxp}. Our calculated $\Upsilon(3S)$    
cross section is consistent with this limit.  While the theoretical approach presented in this Letter  is  applicable at large
transverse momenta, we observe in Fig.~\ref{U276} that the nuclear modification
factor is approximately constant. This allows us to compare in Fig.~\ref{U276cent}  the centrality dependence
in the lowest $p_T=5$~GeV bin,
not very different form the mean $p_T$ of bottomonia at the LHC, to the experimentally measured
$\Upsilon(1S)$ and $\Upsilon(2S)$  $R_{AA}$ dependence  on the number of participants~\cite{Khachatryan:2016xxp}.
Very good agreement between data and theory is observed.

\begin{figure}[!t]
\vspace*{.0in}
\includegraphics[width=3.38in,height=3.5in]{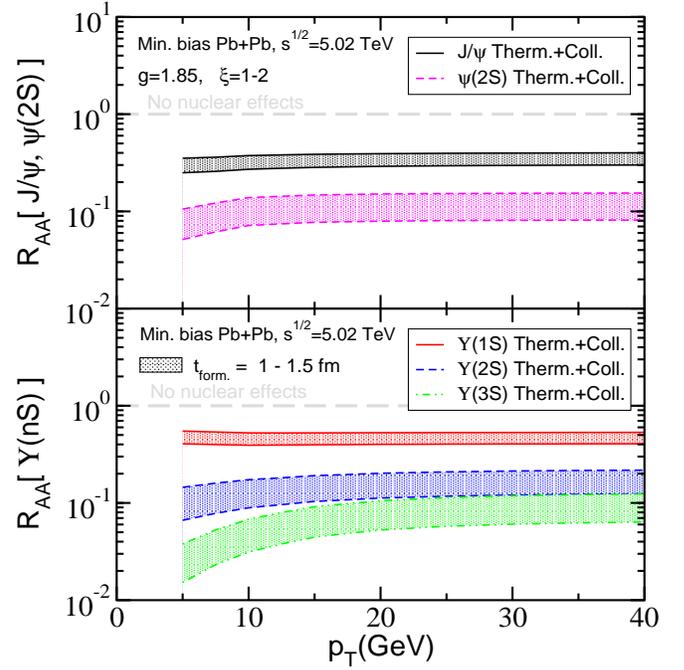} 
\caption{(Color online) Same as in Fig.~\ref{502cen}, but for minimum bias collisions.}
\label{502mb}
\end{figure}

In Figs.~\ref{502cen} and~\ref{502mb}, we present theoretical predictions for
the $R_{AA}$ of various quarkonium species as a function of $p_T$ in Pb+Pb
collisions at $\sqrt{s} = 5.02$~TeV.  The top and bottom panels display results
for charmonium and bottomonium states, respectively. We find a clear separation
in suppression based on how tightly bound the quarkonium state is. We also find
a flat or slightly increasing $R_{AA}$ with $p_T$. By comparing
Fig.~\ref{502cen} to Fig.~\ref{502mb}, we observe that the attenuation of
quarkonia in minimum bias collisions is only slightly smaller than in the most
central collisions. The reason for that behavior is that minimum bias
collisions are strongly dominated by the first $3$ most central classes, as given
by the weights $W_i$ in Eq.~(\ref{cent}). 

\begin{figure}[!t]
\includegraphics[width=3.38in,height=2.2in]{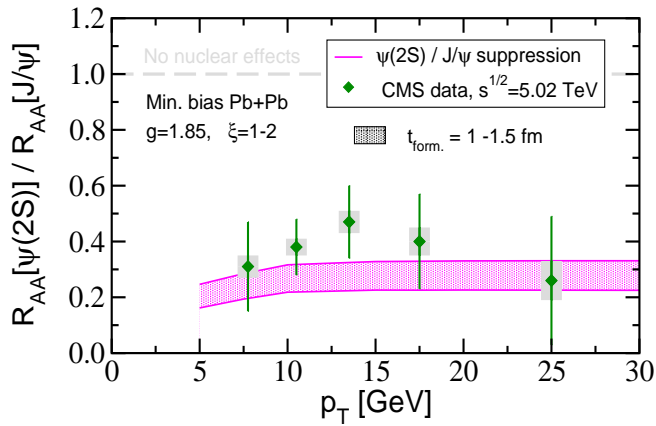} 
\caption{(Color online) Theoretical model predictions for the
  double $(\psi(2S)_{AA}/(\psi(2S)_{pp})/ (J/\psi_{AA}/(J/\psi_{pp})$ ratio in  minimum bias
  Pb+Pb collisions 
at  $\sqrt{S}=2.76$~TeV at the LHC. Data is from CMS~\cite{Sirunyan:2016znt}. }
\label{J2toJ1}
\end{figure}

Very recently, at $\sqrt{s} =5.02$~TeV,  measurements of relative suppression
ratios of excited to ground quarkonium states have
appeared~\cite{Sirunyan:2016znt,Sirunyan:2017lzi}. The data for $\psi(2S) / J/\psi $
is publicly available and shown in Fig.~\ref{J2toJ1}.  Theoretically, the double suppression
ratio can be obtained from the results in Fig.~\ref{502mb} and is compatible
with the experimental data within the statistical and systematic error bars. \\

\section{Conclusions~\label{section:Conclusions}}

In summary, we presented theoretical results for the $p_T$-differential
suppression of charmonia and bottomonia in Pb+Pb collisions at the LHC. The
dynamics of $Q\bar{Q}$ pairs, which evolve into the observed quarkonium states,
 is governed in HICs by the formation and dissociation time scales. A key
element of our formalism that addresses this dynamics is that the formation time
of proto-quarkonia is $\sim 1$~fm. We assume that this time scale is long enough that
the $Q\bar{Q}$ interact via a color screened thermal
potential~\cite{Mocsy:2007jz,Kaczmarek:2005ui}. Therefore we employ the
wavefunctions obtained by solving the Schr\"{o}dinger equation for $Q\bar{Q}$
interacting via a screened potential to calculate the dissociation time scale,
using the theoretical setup described in~\cite{Sharma:2012dy}. The technical
advances that are further incorporated in the calculation are better
constraints on the NRQCD matrix elements that are relevant for the production
of high-$p_T$ $\psi(2S)$ and $\chi_c$ states, and their feed-down  to $J/\psi$,
and 2+1 dimensional event-by-event hydrodynamic modeling of the QGP
background~\cite{Shen:2014vra}. 

We explored the phenomenological implications of this theoretical model for
quarkonium production and propagation in the QGP background created in heavy
ion collisions, first around mid-rapidity in Pb+Pb collisions at the LHC. We
found good separation in the magnitude of the suppression between the ground
and excited charmonium and bottomonium states, compatible with recent
experimental measurements. Our results indicate that effects of thermal
screening of the  $Q\bar{Q}$ attractive potential fully dominate the
attenuation of $\psi(2S)$, $\Upsilon(2S)$ and $\Upsilon(3S)$. On the other
hand, $J/\psi$ and $\Upsilon(1S)$ are also sensitive to the dissociation
processes due to collisional interactions. The approximately constant or
slightly decreasing $R_{AA}$ with $p_T$  predicted by this model arises from
the early ${\cal O}(1{\rm \ fm})$ formation of the interacting quarkonium
state.  The uncertainty of the phenomenological results was estimated by
varying the formation time and the strength of the collisional broadening of
the $Q\bar{Q}$ pair.  We found that the charmonium suppression measurements are
better described by the upper edge of the $R_{AA}$ uncertainty band, whereas
bottomonium suppression measurements are better described by its lower edge.
While an illustrative subset of results was presented here, detailed
predictions are available that will allow to test this model versus upcoming
experimental measurements of quarkonium suppression at $\sqrt{s}=5.02$~TeV. 

In the future, we plan to address data at finite rapidity with the same parameters
and test the model further. This will likely require inclusion of CNM effects since
$p-$Pb data at finite rapidity seems to show non-trivial nuclear modification patterns.
It will also be interesting and instructive to investigate non-prompt $J/\psi$
production, which probes the complementary but different physics of in-medium
modification of heavy-quark parton showers~\cite{Kang:2016ofv,Huang:2013vaa}. \\

{\bf Acknowledgments:} This research is supported by  the US Department of Energy, Office of
Science, under Contract No.~DE-AC52-06NA25396 and in part by the DOE 
Early Career Program. We are grateful to C. Shen for help with the iEBE-VISHNU code.
RS thanks J. P. Blaizot, S. Datta, R. Gavai, S. Gupta, and 
A. Tiwari for rewarding discussions on the topic.



\end{document}